\documentclass[pra,aps,twocolumn,notitlepage,superscriptaddress,showpacs,nofootinbib]{revtex4-1}
\usepackage[colorlinks]{hyperref}
\hypersetup{
	colorlinks = true,
	citecolor = {blue},
	linkcolor = {blue},
    urlcolor={blue}
}
\usepackage{verbatim}
\usepackage{dsfont}
\usepackage{graphicx}
\usepackage{bm}
\usepackage[usenames]{color}
\usepackage{color}
\usepackage{amsmath}
\usepackage[10pt]{moresize}
\usepackage[latin1]{inputenc}
\usepackage{latexsym}
\usepackage{amssymb}
\usepackage{booktabs}
\usepackage{multirow}
\usepackage{subfigure}
\usepackage{dcolumn}
\usepackage{mathrsfs}

\newcommand{\ket}[1]{\ensuremath{|#1\rangle}}
\newcommand{\forget}[1]{}
\usepackage{graphicx}

\graphicspath{{FIG/}}

\begin{document}

\title{Remote blind state preparation with weak coherent pulses in field}

\author{Yang-Fan Jiang}

\author{Kejin Wei}

\author{Liang Huang}
\affiliation{Shanghai Branch, National Laboratory for Physical Sciences at Microscale and Department of Modern Physics, University of Science and Technology of China, Shanghai 201315, China}
\affiliation{Synergetic Innovation Center of Quantum Information and Quantum Physics, University of Science and Technology of China, Hefei, Anhui 230026, China}

\author{Ke Xu}
\affiliation{Centre for Quantum Information and Quantum Control (CQIQC), Dept. of Electrical \& Computer Engineering and Dept. of Physics, University of Toronto, Toronto,  Ontario, M5S 3G4, Canada}

\author{Qi-Chao Sun}

\author{Yu-Zhe Zhang}
\affiliation{Shanghai Branch, National Laboratory for Physical Sciences at Microscale and Department of Modern Physics, University of Science and Technology of China, Shanghai 201315, China}
\affiliation{Synergetic Innovation Center of Quantum Information and Quantum Physics, University of Science and Technology of China, Hefei, Anhui 230026, China}

\author{Weijun Zhang}

\author{Hao Li}

\author{Lixing You}

\author{Zhen Wang}
\affiliation{State Key Laboratory of Functional Materials for Informatics, Shanghai Institute of Microsystem and Information Technology, Chinese Academy of Sciences, Shanghai 200050, China}

\author{Hoi-Kwong Lo}
\email{hklo@ece.utoronto.ca}
\affiliation{Centre for Quantum Information and Quantum Control (CQIQC), Dept. of Electrical \& Computer Engineering and Dept. of Physics, University of Toronto, Toronto,  Ontario, M5S 3G4, Canada}

\author{Feihu Xu}
\email{feihu.xu@ustc.edu.cn}
\affiliation{Shanghai Branch, National Laboratory for Physical Sciences at Microscale and Department of Modern Physics, University of Science and Technology of China, Shanghai 201315, China}
\affiliation{Synergetic Innovation Center of Quantum Information and Quantum Physics, University of Science and Technology of China, Hefei, Anhui 230026, China}

\author{Qiang Zhang}
\email{qiangzh@ustc.edu.cn}
\affiliation{Shanghai Branch, National Laboratory for Physical Sciences at Microscale and Department of Modern Physics, University of Science and Technology of China, Shanghai 201315, China}
\affiliation{Synergetic Innovation Center of Quantum Information and Quantum Physics, University of Science and Technology of China, Hefei, Anhui 230026, China}

\author{Jian-Wei Pan}
\email{pan@ustc.edu.cn}
\affiliation{Shanghai Branch, National Laboratory for Physical Sciences at Microscale and Department of Modern Physics, University of Science and Technology of China, Shanghai 201315, China}
\affiliation{Synergetic Innovation Center of Quantum Information and Quantum Physics, University of Science and Technology of China, Hefei, Anhui 230026, China}

\begin{abstract}
Quantum computing has seen tremendous progress in the past years. Due to the implementation complexity and cost, the future path of quantum computation is strongly believed to delegate computational tasks to powerful quantum servers on cloud. Universal blind quantum computing (UBQC) provides the protocol for the secure delegation of arbitrary quantum computations, and it has received significant attention. However, a great challenge in UBQC is how to transmit quantum state over long distance securely and reliably. Here, we solve this challenge by proposing a resource-efficient remote blind qubit preparation (RBQP) protocol with weak coherent pulses for the client to produce, using a compact and low-cost laser. We experimentally verify a key step of RBQP -- quantum non-demolition measurement -- in the field test over 100-km fiber. Our experiment uses a quantum teleportation setup in telecom wavelength and generates $1000$ secure qubits with an average fidelity of $(86.9\pm1.5)\%$, which exceeds the quantum no-cloning fidelity of equatorial qubit states. The results prove the feasibility of UBQC over long distances, and thus serving as a key milestone towards secure cloud quantum computing.
\end{abstract}

\maketitle

As physicist Richard Feynman realized three decades ago \cite{Feynman1982}, quantum computation holds the promise of exponential speed up over classical computers in solving certain computational tasks. Quantum computation has been an area of wide interest and growth in the past couple of years \cite{harrow2017quantum,mohseni2017commercialize}. Because of implementation complexity, it is speculated that the future quantum computers are accessed via the cloud service for common users. Indeed, the recent effort on quantum cloud service \cite{cloudQC} demonstrates the path towards this speculation. Blind quantum computing (BQC) \cite{Childs2001,ArrighiBQC,BroadbentUBQC} is an effective method for a common user (namely the Client), who has limited or no quantum computational power, to delegate computation to an untrusted quantum organization (namely the Server), without leaking any information about the user's input and computational task.

\begin{figure}
	\centering
	\includegraphics[width=0.4\textwidth]{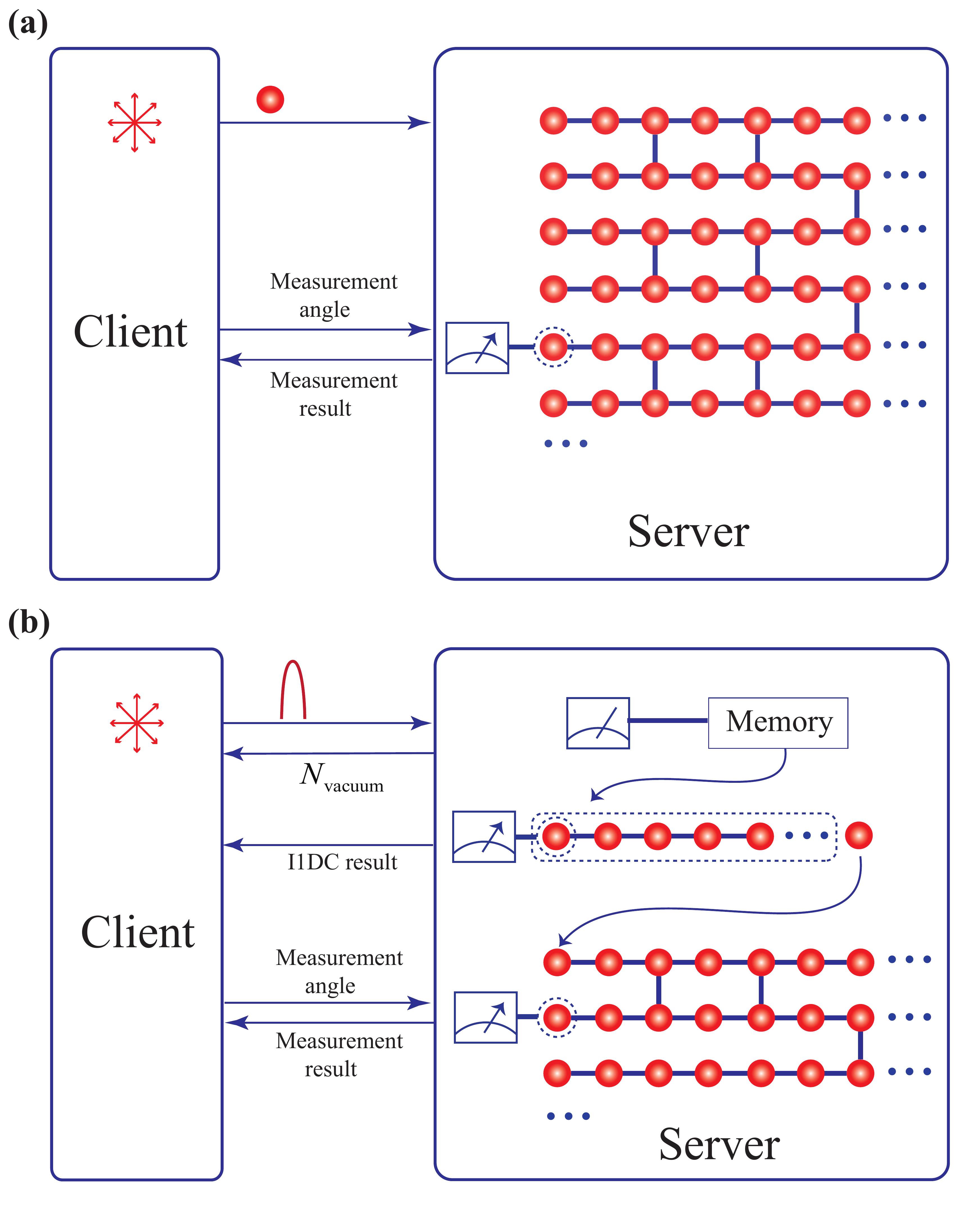}
	\caption{\textbf{(a)}, UBQC with single photons \cite{BroadbentUBQC}.  Client prepares $S$ single qubits randomly prerotated in the polarization states $\ket{+_{\theta_i}}_{i=1}^S=\frac{1}{\sqrt{2}}(\ket{0}+e^{i\theta_i}\ket{1})$, and sends them to Server, who builds up the brickwork state to realize the measurement-based quantum computing. Client transmits measurement angle $\sigma_i =$ ($\phi_i+\theta_i+ r_{i}\pi$ mod $2\pi$) to Server through a classical channel with $r_i\in\{0, 1\}$. Server reports each measurement outcome to Client who performs bit flips if $r_i = 1$. \textbf{(b)}, UBQC with WCPs \cite{Dunjko2012b}. Client prepares a sequence of $N$ phase-randomized WCPs with random polarization $\ket{+_{\theta_i}}_{i=1}^S$, and sends them to Server. Server performs QND measurement on each WCP, stores the non-vacuum pulses and reports the number of vacuum events $N_{0}$ to Client.  Client checks $N_{0}$ and decides whether to continue. If the protocol continues, Server performs the I1DC subroutine on the stored photons and tells Client the results Sever ends up with a perfect random qubit in the state $\ket{+_{\theta}}$, which only Client knows $\theta$. The rest computational steps are the same as \textbf{(a)}. }
	\label{fig:Bqc}
\end{figure}

Various BQC protocols have been proposed in theory \cite{morimae2012blind,Giovannetti2013,Mantri2013,Reichardt2013,fitzsimons2017unconditionally,aharonov2017interactive}. In addition, several experiments have been reported to demonstrate the feasibility of BQC with photonic qubits \cite{Barz2012,Barz2013,Fisher2014,Greganti2016,Gehring2016,Huang2017}. See Ref.~\cite{fitzsimons2017private} for a review. Notably, the universal BQC (UBQC) \cite{BroadbentUBQC} (see Fig.~\ref{fig:Bqc}(a)), built upon the model of measurement-based quantum computation \cite{Raussendorf2001}, does not require any quantum computational power or quantum memory for Client. The security or blindness of the UBQC protocol is information-theoretic, i.e., Server cannot learn anything about Client's computation except its size. The only non-classical requirement for Client is that she can prepare qubits with a single photon source perfectly. Nonetheless, practical single photon sources are not yet readily available, despite a lot of effort \cite{aharonovich2016solid}.

To resolve the state-preparation issue, the recent remote blind qubit preparation (RBQP) protocol, proposed in \cite{Dunjko2012b}, enables preparing blind qubits with weak coherent pulses (WCPs), generated from a compact and low-cost laser diode, instead of perfect single photon source. In this protocol, Client prepares a sequence of WCPs with random polarization ${\theta _i}{ \in _{\mathop{\rm R}\nolimits} }\{ {{k\pi } \mathord{\left/{\vphantom {{k\pi } 4}} \right. \kern-\nulldelimiterspace} 4}:0 \le k \le 7\}$ and sends them to Server through a quantum channel. Server performs quantum non-demolition~(QND) measurements on each of received WCPs and declares the results to Client. Client checks the reported number of vacuum events: if the number is smaller than a preset threshold, she asks Server to perform the interlaced 1-D cluster computation~(I1DC) subroutine \cite{Dunjko2012b} on the non-vacuum pulses. The RBQP protocol is completed with a polarization angle $\theta$ which is only known by Client and a \emph{single} qubit in the state $\ket{+_{\theta}}$ held by Server. Running the RBQP protocol $S$ times will result in a computational size of $S$ single qubits. For a channel with transmittance $\eta$, this requires a total number of $N$ WCPs \cite{Dunjko2012b},
\begin{eqnarray}
N \ge \frac{{18\log ({S \mathord{\left/
				{\vphantom {S \varepsilon }} \right.
				\kern-\nulldelimiterspace} \epsilon })}}{\eta^4},\label{original}
\end{eqnarray}
where $\epsilon$ denotes the failure probability. Nonetheless, the RBQP is inefficient for small $\eta$, i.e., $N$ scales as ${\rm O}({1 \mathord{\left/{\vphantom {1 {{\eta ^4}}}} \right.\kern-\nulldelimiterspace} {{\eta ^4}}})$. It is thus demanding to design an efficient protocol for the future quantum network, where Client can access Server over a long distance.

We propose a refined RBQP protocol by employing the decoy state method, which is originally invented in the field of quantum key distribution \cite{Lo2005,wang2005beating}. Our protocol can greatly reduce the required number of WCPs from ${\rm O}({1 \mathord{\left/
		{\vphantom {1 {{\eta ^4}}}} \right.
		\kern-\nulldelimiterspace} {{\eta ^4}}})$ to ${\rm O}({1 \mathord{\left/
		{\vphantom {1 \eta }} \right.
		\kern-\nulldelimiterspace} \eta })$. Furthermore, instead of generating one single qubit in each run, our protocol allows a client to generate S qubits simultaneously in a single instance. In our protocol, Client randomly modulates the intensity of each WCP according to intensity choice $\mu$ (signal), $\nu$ (decoy) and $0$ (vacuum). Client runs the same as the initial RBQP, but with a different post-processing. With the reported QND results for each intensity, Client performs the decoy-state analysis to estimate the lower bound of the number of single-photon events \cite{Lo2005,wang2005beating}. If the bound is larger than her preset threshold, Client asks Server to discard all the decoy pulses and randomly divided the remaining $M_\mu$ signal pulses into $S$ groups, each group containing $m =M_\mu/S$ signal pulses. Server performs the I1DC subroutine \cite{Dunjko2012b} on each group and returns the measurement results to Client. The protocol completes with $S$ single qubits held by Server, of which the polarization angles are only known to Client. By doing so, in the limit that the probability of sending a signal state  is approximately 1, the lower bound of $N$ in our protocol is,
\begin{equation}\label{Decoy}
N \ge \frac{{2.1S\log ({S \mathord{\left/
				{\vphantom {S \varepsilon }} \right.
				\kern-\nulldelimiterspace} \epsilon })}}{\eta }.
\end{equation}

Comparing with Eq.~\eqref{original}, $N$ scales as ${\rm O}({1 \mathord{\left/
		{\vphantom {1 \eta }} \right.
		\kern-\nulldelimiterspace} \eta })$, which is far less than that of the original protocol. We remark that any failure to detect a photon is subjected to the loss, which does not affect the security. We have also derived the analysis after considering the finite-data effect and show the details of these results in Appendix~\ref{finite}.

\begin{figure*}
	\centering
	\includegraphics[width=0.75\textwidth]{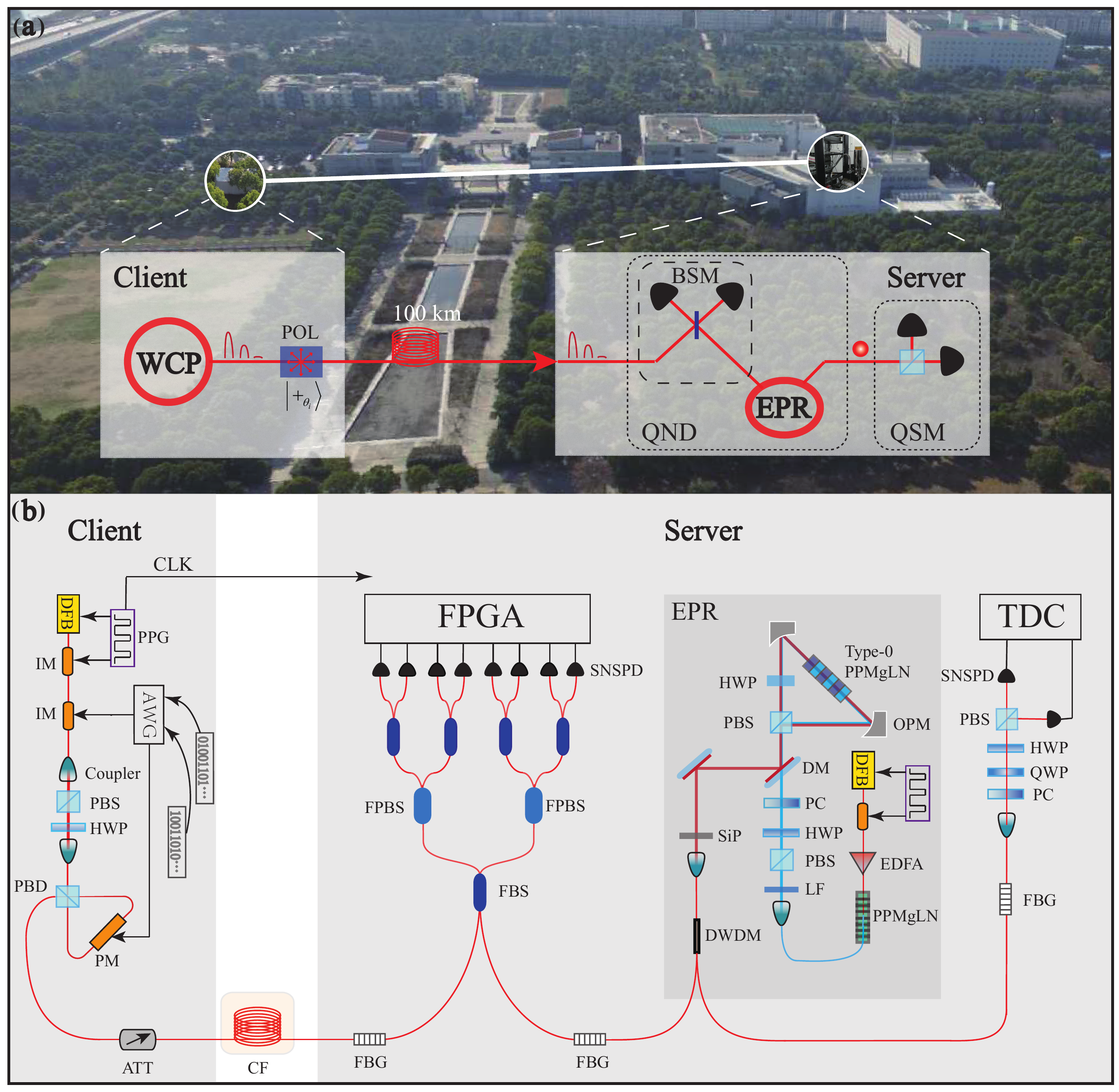}
	\caption{\textbf{(a)}, Birds-eye view of the experiment between Client and Server over a field distance of 199 m. Client sends WCPs, in polarization states of $\ket{+_{\theta_i}}$ with signal and decoy intensities, to Server who implements QND measurement based on quantum-teleportation and quantum-state-tomography measurements (QSM). \textbf{(b)}, Experimental setup. Client's setup: Client generates laser pulses using a distributed feedback~(DFB) laser and an intensity modulator~(IM), which are driven by a pulse pattern generator~(PPG). The other IM is used to generate signal and decoy intensity randomly. The states of $\ket{+_{\theta_i}}$ are encoded into the pulse by utilizing a loop-interferometer-based polarization modulation, which consists of a polarization beam displacer~(PBD) and a phase modulation~(PM). All the encodings are controlled by an arbitrary waveform generator (AWG) with independent random numbers. The pulses are  attenuated by an attenuator and sent to Server through a standard coiled fiber. Server's setup: the laser pulses from an 1558~nm gain-switched DFB are amplified by an erbium doped fiber amplifier (EDFA) and up-converted to 779~nm pulses in an in-line periodically poled MgO doped Lithium Niobate~(PPMgLN) crystal. The produced 779~nm pulses are focused into the second PPMgLN in the Sagnac loop to generate polarization-entangled photon pairs. The signal and idler photons are singled out by an inline dense wavelength division multiplexing filters~(DWDM); one is used to implement the Bell state measurement (BSM) and the other is used to perform QSM. The implementation of QSM includes a polarizing beam splitter~(PBS), two superconducting nanowire single-photon detectors~(SNSPDs) and a time-to-digital converter (TDC). CLK: synchronization signal; FBG: fiber Bragg grating; FBS, fiber beam splitter; FPBS, fiber polarizing beam splitter; FPGA, field programmable gate array; HWP, half wave plate;  LF, low-pass filter; PC, phase compensator; OPM, off-axis parabolic mirror; DM, dichroic mirror; SiP, silicon pellet.}
	\label{fig:setup}
\end{figure*}

A key challenge to implement RBQP is the realization of QND measurement. QND is a crucial technology in quantum information and it has been investigated widely in matter-based platforms \cite{guerlin2007progressive,reiserer2013nondestructive}. However, these matter-based realizations require challenging techniques, such as strong light-matter interactions and optical wavelength conversion, which are not mature for real-life applications. Here, we solve the challenge by designing an experimentally feasible scheme based on linear optics and teleportation-based method  \cite{jacobs2002quantum,wang2015quantum,HirokiOpticaTele,Sun2016,Valivarthi2016a}. We move the QND to the field test over 100-km fiber by using two independent photon sources. The scheme of our experiment is shown in Fig.~\ref{fig:setup}(a). We construct a quantum link in the field at the city of Shanghai, in which Client sends the polarization-encoding~(POL) WCPs with decoy states to Server who performs QND measurements. The field distance between Client and Server is about 199~m.

Fig.~\ref{fig:setup}(b) shows details of our experimental realization. Client possesses a gain-switched distributed feedback laser~(DFB) to generate laser pulses at a repetition frequency of 250~MHz. Each pulse is carved into 37~ps pulse duration after passing through the first intensity modulator~(IM). To generate the two decoy states, intensities of the pulses are randomly modulated by the second IM. Key bits are encoded into polarization states of the WCPs by a loop--interferometer--based polarization encoding scheme which consists of a polarization beam displacer~(PBD) and phase modulator~(PM). After attenuation, Client sends the weak coherent pulses to Server through a standard telecom coiled fiber.

Server prepares Einstein-Podolsky-Rosen (EPR) pairs of signal~(s) and idler~(i) photons in the quantum state of $\ket{\Phi^+}_{si}=\frac{1}{\sqrt{2}}(\ket{H}_s\ket{H}_i+\ket{V}_s\ket{V}_i)$ via spontaneous parametric down-conversion~(SPDC) process. The signal and idler photons are singled out by an inline dense wavelength division multiplexing filter (DWDM). The signal photons are used to take a Bell state measurement with the received photons from Client. These photons are detected by  high-quality superconducting nanowire single-photon detectors~(SNSPDs), where the detection events are registered by a field programmable gate array~(FPGA). Note that after fiber polarization beam splitters~(FPBSs), we use four fiber beam splitters~(FBSs) and eight SNSPDs to mimic photon-number-resolving detectors \cite{divochiy2008superconducting}. This allows us to probabilistically detect 2-or-more inbound photons from the WCP. The idler photons undergo a quantum state tomography measurement for the quantification of the quality of the prepared qubits.

To implement the protocol, there are several technical challenges. First, a high-speed and high-fidelity polarization modulation is required to prepare eight polarization states $\theta _i$. We use a loop-interferometer-based scheme to realize the polarization modulation at a rate of 250~MHz with an average fidelity of ($99.42\pm0.09)\%$ \cite{agnesi2019all}. Second, it requires a high-visibility interference between two independent sources, i.e., the EPR pairs and the WCPs which experiences a long-distance transmission. To do so, we synchronize the two independent sources with a~12.5 GHz microwave clock and exploit two fiber Bragg gratings (FBG) filters with a bandwidth of~3.3 GHz to suppress the spectral distinguishability. Third, we optimize the average photon number from the WCP to obtain an optimal interference visibility. Finally, we detect the photons with a combination of four FBSs to decrease the multi-photons effect and eight high-efficiency and low-dark-count SNSPDs to maximize the interference visibility. See Appendix~\ref{app:experiment} for further details. These efforts allow us to achieve a high QND measurement fidelity of about $95\%$, which is much higher  than those reported in previous works, e.g., $75\%$ in \cite{Valivarthi2016a}.

\begin{figure}
	\centering
	\includegraphics[width=0.44\textwidth]{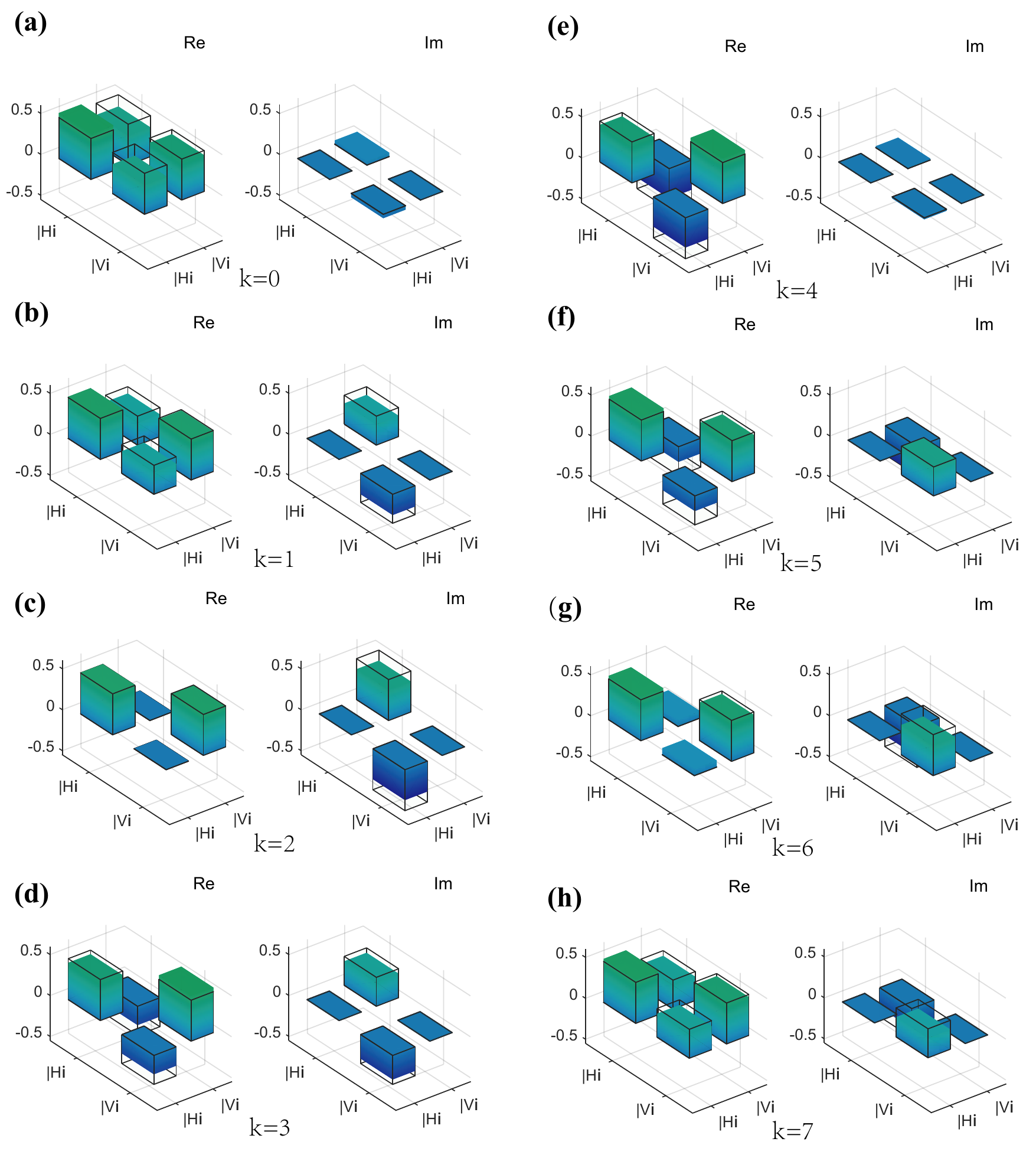}
	\caption{\textbf{(a)-(h),} The real and imaginary parts of the reconstructed density matrices for eight polarization states $\ket{+_{\theta_i}}=\frac{1}{\sqrt{2}}(\ket{0}+e^{i\theta_i}\ket{1})$   with ${\theta_i} \in \{k\pi/4 :0 \le k \le 7\} $ after QND measurement over 100~km fiber. The black frames denote the ideal density matrices.  The average fidelity is characterized as $(86.9\pm1.5)\%$. The error bar represents one standard deviation.}\label{fig:states}
\end{figure}

We characterize the QND test by performing quantum-state-tomography measurements on the teleported quantum states. We run our protocol over a distance of 100~km fiber, and measure the density matrices of eight teleported states at Server. These results are shown in Fig~\ref{fig:states}. The average fidelity is characterized as $(86.9\pm1.5)\%$, which exceeds the maximum value of $2/3$ achievable in classical teleportation, and the quantum phase-covariant no-cloning  bound of  $85.4\%$ \cite{2000Bru,PhysRevLett.94.040505}. This result indicates the high fidelity of our QND measurement.

\begin{figure}
	\centering
	\includegraphics[width=0.4\textwidth]{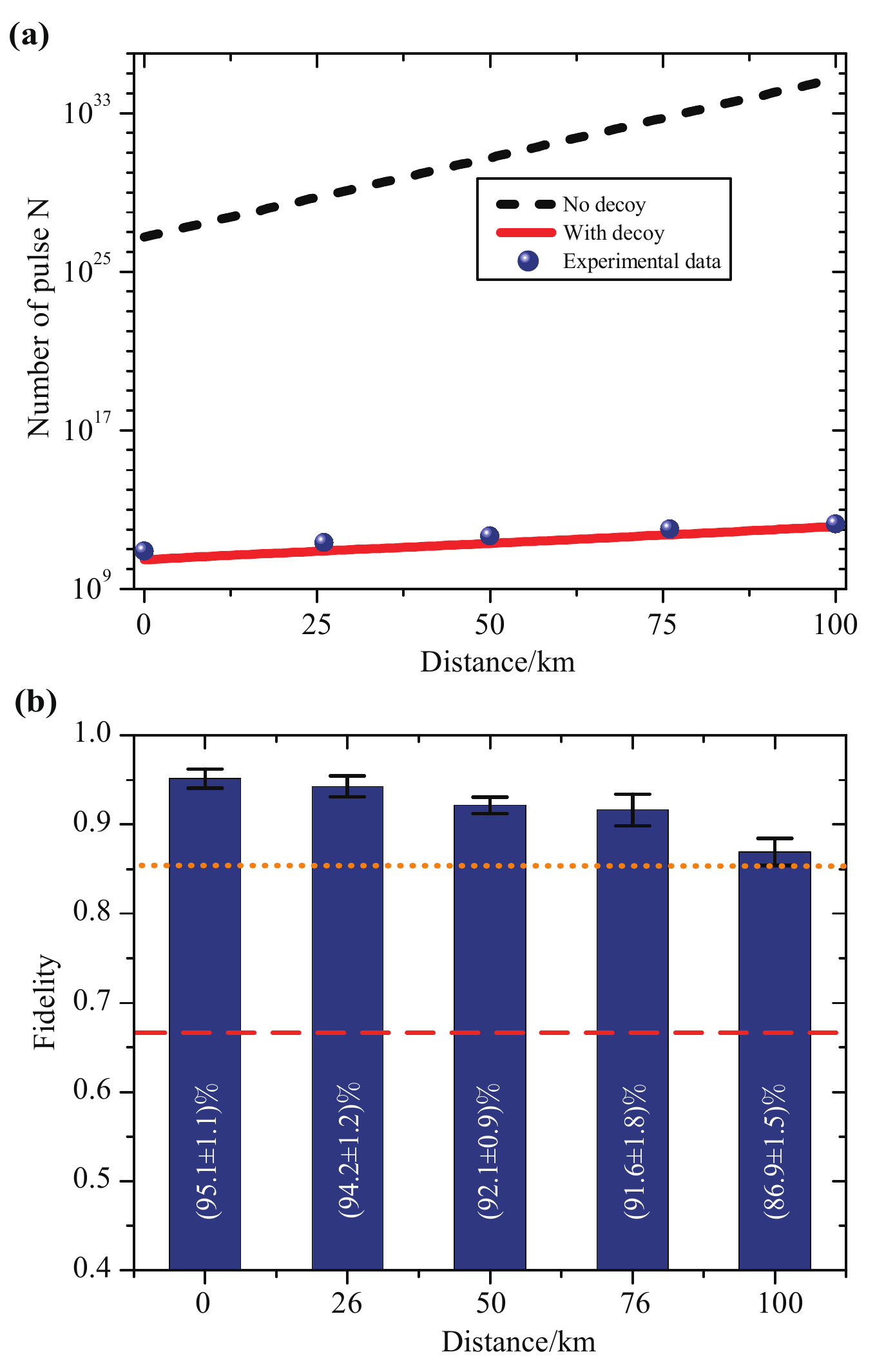}
	\caption{\textbf{(a),} The required number $N$ of WCPs for preparing $1000$ secure qubits. The dashed black curve and solid red curve are numerical simulation of $N$ for RBQC with and without decoy states \cite{Supplemental}. The blue dots are our experiment results. \textbf{(b),}  The average fidelities of the polarization states after QND measurement. The fidelities are measured using quantum state tomography. The error bars represent one standard deviation. All fidelities exceed both the classical fidelity limit of 2/3, represented by the dashed-red line, and the  quantum phase-covariant no-cloning  bound of  $85.4\%$, represented by the dot-orange line.}
	\label{fig:result}
	
\end{figure}

We run the whole system with fibers at distances 0~km 26~km, 50~km, 76~km and 100~km. Experimental parameters, including the intensities and probability distributions of signal and decoy pulses, are optimized numerically (see Appendix~\ref{finite}). In each run, we generate $S = 1000$ qubits which could be made blind via the I1DC. The experimental results are shown in Fig.~\ref{fig:result}(a). We can see that the required $N$ of our protocol is much lower than that of the original protocol \cite{Dunjko2012b}. In particular, at the distance of 100~km, it is up to 20 orders of magnitudes lower than that of the original protocol. At 0~km, the loss primarily comes from the inefficient QND measurement. Such a huge effective loss due to an inefficient QND measurement causes that the original RBQP protocol requires at least $N \sim 10^{26}$ pulses. In contrast, our decoy-state based protocol requires only $N \sim 10^{10}$ pulses. This number of pulses can be generated in less than a minute using our implementation system. Even at 100~km distance, our experiment only needs about 2 hours to generate $S = 1000$ blind qubits. The average fidelities of the eight polarization states $\ket{+_{\theta_i}}$ for different distances are shown in Fig.~\ref{fig:result}(b).

In the RBQP, as shown in Fig.~\ref{fig:Bqc}(b), the signal WCPs should be stored in a quantum memory after the QND measurement and the I1DC is applied afterwards. We simulate this procedure by storing the density matrixes of the signal states and performing the I1DC subroutine on a personal computer (see Appendix~\ref{app:I1DC}). Our simulation results show that at the fiber length of 0~km, the average fidelity of the $1000$ blind qubits is $(81.9\pm2.0)\%$. This fidelity can be improved if the client uses error correction code for encoding. A full implementation demands a high-performance quantum memory. In our setup, to generate 1000 blind qubits at 100 km would require a storage time of $\sim$2 hours and near unity process fidelity, which is still beyond the current quantum memory technology. Nevertheless, long storage time, large bandwidth and high fidelity quantum memories have been achieved, recently \cite{zhou2012realization,zhong2015optically,yang2016efficient,jiang2019experimental}. These subjects are important for future studies.

In summary, we have proposed a decoy-state RBQP protocol and reduce the required number of WCPs $N$ from ${\rm O}({1 \mathord{\left/
		{\vphantom {1 {{\eta ^4}}}} \right.
		\kern-\nulldelimiterspace} {{\eta ^4}}})$ to ${\rm O}({1 \mathord{\left/
		{\vphantom {1 \eta }} \right.
		\kern-\nulldelimiterspace} \eta })$ to generate $S$ blind qubits. We have demonstrated a key step of our protocol by implementing the QND with two independent photon sources in the field, up to 100 km fiber. The fidelity of the generated qubits is above $86\%$. Our RBQP protocol with WCP and photonic experiment lead a heuristic exploration for UBQC over long-distance quantum networks, and they will be a crucial step for the commercialization and widespread adoption of secure quantum computation in cloud.

\begin{acknowledgements}
The authors would like to thank Bing Bai, Tong Xiang, Xiaohui Bao and Yong Yu for helpful discussions. This work was supported by National Key R\&D Program of China (2018YFB0504300), the National Natural Science Foundation of China, the Chinese Academy of Science. H.-K. Lo was supported by NSERC, US Office of Naval Research, CFI, ORF, and Huawei Canada.

Y-F. Jiang and K. Wei contributed equally to this work.
\end{acknowledgements}

\newpage

\appendix

\section{Remote blind qubit state preparation with decoy states} \label{finite}
\subsection{Detailed steps}
Here, we show the details of the  two-decoy states method, where, besides the signal state $\mu$, the client prepares two decoy states: weak decoy $\nu$ and vacuum decoy $0$. The protocol goes as follows:

(\romannumeral1) Client prepares $N$ phase-randomized WCPs, in
which $\{ {N_\mu },{N_\nu },{N_0}\} $ are the number of pulses for signal state and decoy states with intensity $\{ \mu ,\nu ,0\}$ and
probability $\{ {p_\mu },{p_\nu },{p_0}\}$. Each pulse is randomly polarized with polarization ${\theta _i}{ \in _{\mathop{\rm R}\nolimits} }\{ {{k\pi } \mathord{\left/
		{\vphantom {{k\pi } 4}} \right.
		\kern-\nulldelimiterspace} 4}:0 \le k \le 7\}$. Client sends the pulses to Server through a quantum
channel with a transmittance no less than $\eta$.

(\romannumeral2) After Server receives the pulses, he performs QND measurements and reports the measurement results to the client. Client performs the decoy states analysis to estimate the lower bound of the number of single-photon events, i.e., Client calculates the gain of signal and decoy states from the reported results of Server and then estimates the lower bound of the gain of single-photon events following the method proposed in~\cite{Ma2005}. Client checks the estimated number of single-photon events: if the number is greater than her predetermined threshold of Eq.~(\ref{threshold}),  she continues; otherwise the protocol aborts.

(\romannumeral3) If the protocol is not aborted, Client then asks Server to discard all the decoy states. Server is
now left with $M_\mu$ signal states he received. Client
then asks the server to randomly divide these $M_\mu$ signal
states into S groups, each group containing $m =M_\mu/S$
signal states. Server performs the I1DC subroutine~\cite{Dunjko2012b} on each group of the signal states
and returns the measurement results to Client. The
protocol is completed with $S$ single qubits held by Server, of which the polarization angles are only known
to Client.
\subsection{Security analysis}

The security (i.e., blindness) of our protocol lies in the fact
that in the I1DC subroutine as long as the server is ignorant
of the polarization angle of at least one photon of the 1D cluster, he is totally ignorant of the polarization angle of the final
qubit~\cite{Dunjko2012b}. Therefore the task of the client is to make sure that there is at least one single photon in each group. We now show how to choose a proper $N$ so that the probability that the protocol fails ${P_f}$ is bounded by ${P_f} \le \epsilon$ for given transmittance $\eta$ and computation size $S$.

Suppose there are $M_\mu ^1$ single photon states in the $M_\mu$ signal states received by Server. We define $p_1$ as the single photon ratio in the signal states ${p_1} \equiv \frac{{M_\mu ^1}}{{{M_\mu }}}$, and $p_f ^1$ as the probability that
one of the groups fails, i.e. there is no single photon in that
group. Then $p_f ^1$ can be estimated by
\begin{equation}
p_f^1 = \frac{{\left( {\begin{array}{c}
			{{M_\mu } - M_\mu ^1}\\
			m
			\end{array}} \right)}}{{\left( {\begin{array}{c}
			{{M_\mu }}\\
			m
			\end{array}} \right)}} \simeq \left( \frac{{M_\mu } - M_\mu ^1}{M_\mu }\right)^m = {(1 - {p_1})^m}.
\end{equation}

The second equation above is due to the assumption that
$M_\mu,~(M_\mu -M_\mu^1) \gg m$. The probability that the protocol fails  $P_f$ now
can be bounded by
\begin{equation}\label{fail}
{P_f} = Sp_f^1 = {S(1 - {p_1})^m} \le \epsilon
\end{equation}

Hence we have the bound for the number of single-photon
events
\begin{equation}
M_\mu ^1 \ge {M_\mu }[1 - \exp ({\textstyle{{S\log ({\epsilon  \mathord{\left/
					{\vphantom {\epsilon  S}} \right.
					\kern-\nulldelimiterspace} S})} \over {{M_\mu }}}})].
\end{equation}
Also we can derive the required number of signals $N$. From Eq.~(\ref{fail}), we have
\begin{equation}
m > \frac{{\log ({\epsilon  \mathord{\left/
				{\vphantom {\epsilon  S}} \right.
				\kern-\nulldelimiterspace} S})}}{{\log (1 - {p_1})}}.
\end{equation}
Because ${M_\mu } = N{p_\mu }{Q_\mu }$, we obtain the lower bound of
\begin{equation}
N = \frac{{{M_\mu }}}{{{p_\mu }{Q_\mu }}} > \frac{S}{{{p_\mu }{Q_\mu }}} \cdot \frac{{\log ({\epsilon  \mathord{\left/
				{\vphantom {\epsilon  S}} \right.
				\kern-\nulldelimiterspace} S})}}{{\log (1 - {p_1})}}.
\end{equation}
Here, $Q_\mu$ denotes the gain of signal states, which is estimated by Client from Server's feedback.

To evaluate the lower bound of $N$, the key point is lower bound $p_1$. We achieve this using decoy state method. If the Client can prepare infinite decoy states, the client can estimate the single-photon events perfectly. For long distance, i.e., $\eta \ll 1$, we have
\begin{eqnarray}
\begin{aligned}
&{Q_\mu } = 1 - {e^{ - \eta \mu }} \simeq \eta \mu, \\
&Q_1^\mu \simeq \eta \mu {e^{ - \mu }}.
\end{aligned}
\end{eqnarray}

Now $N$ is given by
\begin{eqnarray}
N > \frac{S}{{{p_\mu }\eta \mu }} \cdot \frac{{\log ({\epsilon  \mathord{\left/
				{\vphantom {\varepsilon  S}} \right.
				\kern-\nulldelimiterspace} S})}}{{\log (1 - {e^{ - \mu }})}}
\end{eqnarray}
By solving $\frac{{\partial N}}{{\partial \mu }} = 0$, we obtain the optimal value of $\mu$ to be
$\mu_{opt}=0.7$. In the limit that $p_\mu \simeq 1$, we have the bound for $N$ as
\begin{equation}
N > \frac{{2.1S\log ({S \mathord{\left/
				{\vphantom {S \varepsilon }} \right.
				\kern-\nulldelimiterspace} \epsilon })}}{\eta }
\end{equation}

However, in practice, the client has finite resources and all real-life experiments are done in a finite time. This means that we would use finite decoy states with considering finite statistics. Here, we achieve this using two decoy-state method proposed in \cite{Ma2005}. The lower bound of single-photon events is given by
\begin{eqnarray}
\begin{aligned}
M_\mu ^1 &\ge M_\mu ^{1,L}\\ =
&\frac{{N{p_\mu }{\mu ^2}{e^{ - \mu }}}}{{\mu \nu  - {\nu ^2}}}[Q_\nu ^ - {e^\nu } - Q_0^ +  - \frac{{{\nu ^2}}}{{{\mu ^2}}}(Q_\mu ^ + {e^\mu } - Q_0^ - )],\label{threshold}
\end{aligned}
\end{eqnarray}
where $Q_\lambda ^ + $ and $Q_\lambda ^ - $  denotes respectively the upper bound and the lower bound for the gain of an intensity choice $\lambda  \in \{ \mu ,\nu ,0\}$ due to finite statistics. They are bounded by Hoeffding inequality:
\begin{align}
Q_\lambda ^ \pm  = {Q_\lambda } \pm \sqrt {{Q_\lambda }{{\log ({1 \mathord{\left/
					{\vphantom {1 {{\epsilon _d})}}} \right.
					\kern-\nulldelimiterspace} {{\epsilon _d})}}} \mathord{\left/
			{\vphantom {{\log ({1 \mathord{\left/
								{\vphantom {1 {{\varepsilon _d})}}} \right.
								\kern-\nulldelimiterspace} {{\epsilon _d})}}} {(2{N_\lambda })}}} \right.
			\kern-\nulldelimiterspace} {(2{N_\lambda })}}}
\end{align}
where ${{\epsilon _d}}$ is the failure probability in decoy-state analysis.

The lower bound of $N$ is
\begin{eqnarray}
N > \frac{S}{{{p_\mu }Q_\mu ^ + }} \cdot \frac{{\log ({S  \mathord{\left/
				{\vphantom {\varepsilon  {S)}}} \right.
				\kern-\nulldelimiterspace} {\epsilon)}}}}{{(1 - p_1^ - )}},
\end{eqnarray}
where $p_1^-=M_\mu^{1,L}/NQ_\mu^+$.

\subsection{Numerical simulation and optimization} We perform a simulation based on the experimental parameters of our setup listed in Table \ref{parameters}.
\begin{table}	\caption{Experimental parameters for simulation. $\epsilon$, failure probability; $\epsilon_d$, decoy failure probability; $S$, size; $\eta_d$, detector system efficiency; $P_{dark}$, detector dark count rate; $\mu_{SPDC}$, SPDC's mean photon pair per pulse; $\alpha$, loss coefficient of optical fiber. }
	\centering
	\begin{tabular}{ccccccc}
		\hline\hline
		$\epsilon$&$\epsilon_d$&$S$&$\eta_d$&$P_{dark}$&$\mu_{SPDC}$&$\alpha$\\
		\hline
		$10^{-10}$&$10^{-10}$&$10^3$&$0.105$&$4 \times 10^{-7}$&$0.002$&$0.2~dB/km$\\
		\hline\hline
	\end{tabular}\label{parameters}

\end{table}
To find the optimal values $\{\mu,~\nu,~p_\nu,~p_\mu\}$ for given $\eta$, $S$, and $\epsilon$. In principle, we can done by solving
\begin{eqnarray}
\frac{{\partial N}}{{\partial \mu }} = 0,~ \frac{{\partial N}}{{\partial \nu }} = 0,~\frac{{\partial N}}{{\partial p_\mu }} = 0,~\frac{{\partial N}}{{\partial p_\nu }} = 0,
\end{eqnarray}
which is a complicatedly mathematical problem. Instead, we do it by numerically solving the following routine:
\begin{eqnarray}
\begin{aligned}
&\textbf{min :} ~N,\\
&\textbf{s.t. :} ~S{(1 - \frac{{M_\mu ^{1,L}}}{{M_\mu ^ + }})^{\frac{{M_\mu ^ - }}{S}}} \le \epsilon.
\end{aligned}
\end{eqnarray}
Here, $M_u^{ \pm }$ represent the upper and lower bound of signal states due to finite effects, which are given by $M_\mu ^ \pm  = {M_\mu } \pm \sqrt {{M_\mu } \cdot \log (1/{\varepsilon _d})/2} $.

\section{Experimental details} \label{app:experiment}
\subsection{Single photon polarization state modulation}
\begin{figure}
\centering
\includegraphics[width=0.45\textwidth]{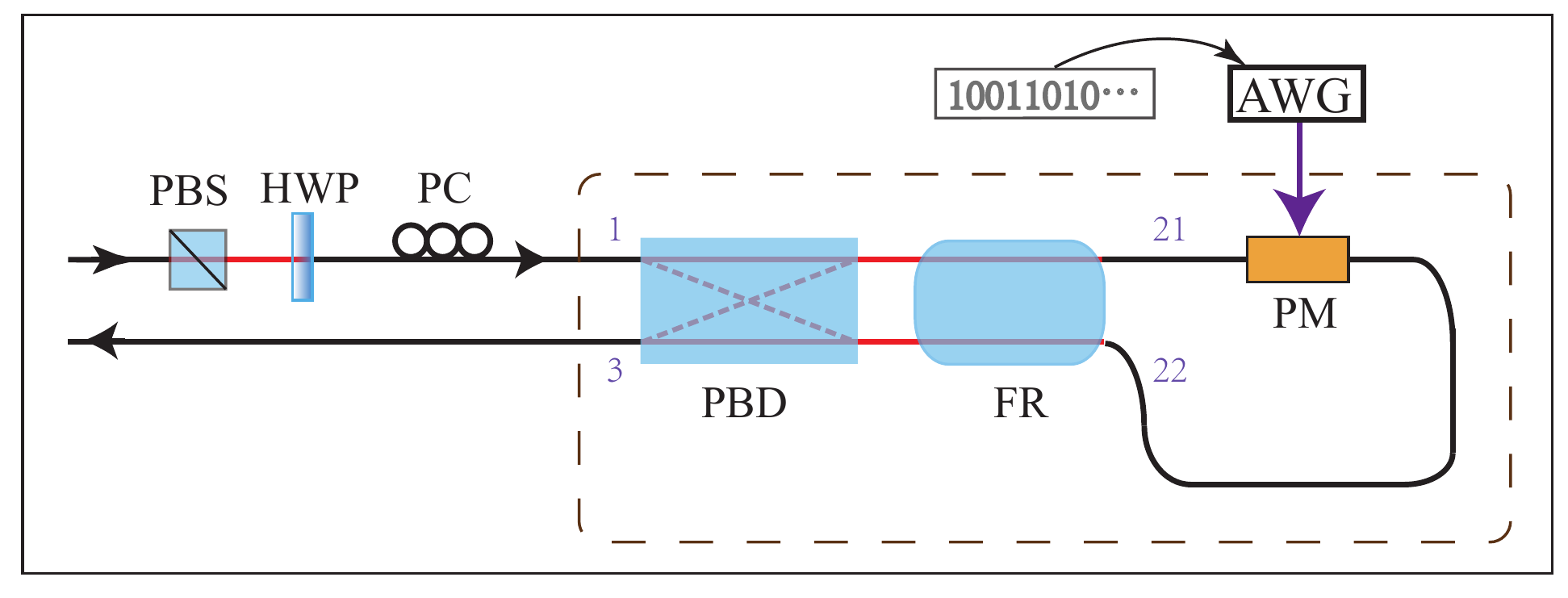}
\caption{The schematic of polarization modulation. The red line denotes space path, the black line denotes fiber. PBS, polarization beam spliter; HWP, half wave plate; PC, polarization controller; PBD, polarization beam displacer; FR, Faraday mirror; PM, phase modulator; AWG, arbitrary waveform generator.}
\label{fig:PM2}
\end{figure}

\begin{figure}
\centering
\includegraphics[width=0.48\textwidth]{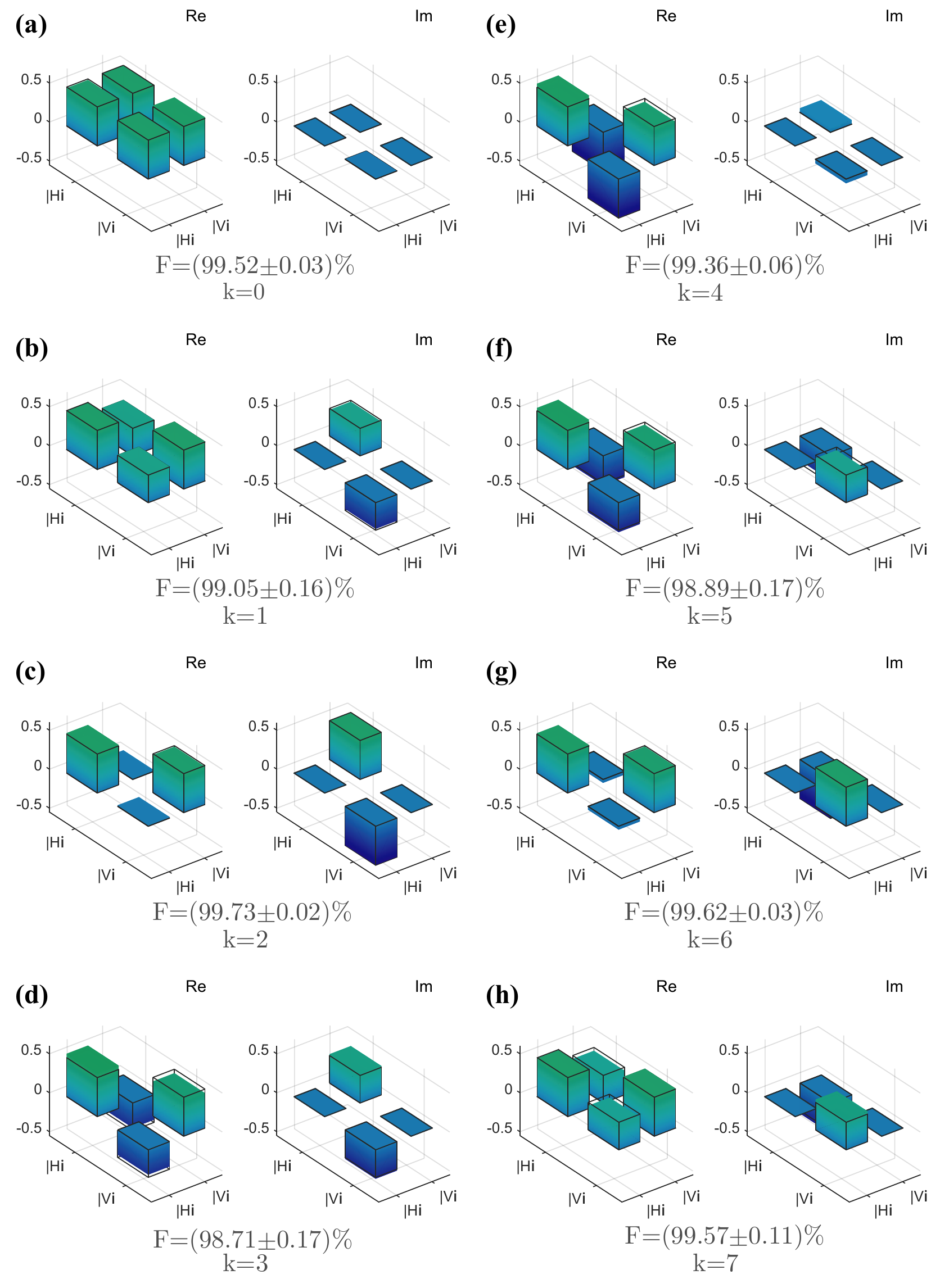}
\caption{\textbf{(a)-(h),} The real and imaginary parts of the reconstructed density matrices for eight polarization states. The black frames denote the ideal density matrices.}
\label{fig:prep}
\end{figure}

We use a loop-interferometer-based polarization encoding scheme, shown in Fig.~\ref{fig:PM2}, to achieve states $\ket{+_{\theta_i}}=\frac{1}{\sqrt{2}}(\ket{H}+e^{i\theta_i}\ket{V})$ at a rate of 250 MHz with high-fidelity. The photons on the state of $\ket{+}$, prepared by a polarization beam spliter (PBS), a half wave plate and polarization controller, incident into a loop interferometer via the port 1 of a polarization beam displacer~(PBD), and then is split to two orthogonal components~(i.e., horizontal~(H) and vertical~(V)). The two polarization components are coupled into the slow axis of the polarization maintained fiber pigtails of the phase modulator (PM) . The PM manipulates eight phases $\theta_i$ to the vertical component randomly via eight random voltages amplitudes generated by a 25~GHz arbitrary waveform generator~(AWG). After routing the PM, the polarization of the two components is exchanged via an Faraday rotator (FR). Hence, when recombined on the PBD and the photon states are output from the port 3. We test the performance of our scheme by reconstructed the eight states using quantum state tomography measurement \cite{DanielPRATomo}. As shown in Fig.~\ref{fig:prep}, we realize each state with high fidelity, and the average fidelity is up to $(99.42\pm 0.09)\%$.

\subsection{Polarization Einstein-Podolsky-Rosen sources}
As depicted in the figure 2 of the main text, we generate polarization Einstein-Podolsky-Rosen~(EPR) pairs in the Bell state $\ket{\Phi^+}_{si}=\frac{1}{\sqrt{2}}(\ket{H}_s\ket{H}_i+\ket{V}_s\ket{V}_i)$ via spontaneous parametric down-conversion~(SPDC) process in a Sagnac loop, here $s$ denotes signal photon and $i$ denotes idler photon. An 1558~nm gain-switched distributed feed-back laser~(DFB) emits 2~ns laser pulse at 250~MHz. All the laser pulses are generated from vacuum fluctuation, so the source is phase-independent. A 40~GHz intensity modulation~(IM) modulates the pulses into 80~ps laser pulses. Both the DFB laser and the IM are driven by a pules pattern generator~(PPG). The laser pulses are amplified by an erbium-doped fibre amplifier~(EDFA) and frequency-doubled in a periodically poled MgO doped Lithium Niobate~(PPMgLN) crystal. We remote the remanent 1558~nm pulse with a low-pass filter~(LF). The 779~nm pump laser is focused into a 2.5~cm long type-0 PPMgLN crystal to generate polarization-entangled photon pairs, with the beam waist of 54~um by using an aspheric lens and an off-axis parabolic mirror~(OPM). The polarized photon pairs are non-degenerated at 1556~nm and 1560~nm, and are coupled into a single-mode optical fibre for spatial mode cleaning. The pump laser is removed by a silicon pellet~(SiP). We create the entangled photon-pair source of $\ket{\Phi^+}_{si}=\frac{1}{\sqrt{2}}(\ket{H}_s\ket{H}_i+\ket{V}_s\ket{V}_i)$  by adjusting the 780~nm half wave plate~((HWP) and the phase compensator~(PC). The signal and idler photons are singled out by inline dense wavelength division multiplexing filters~(DWDM).

\begin{figure}
\centering
\includegraphics[width=0.48\textwidth]{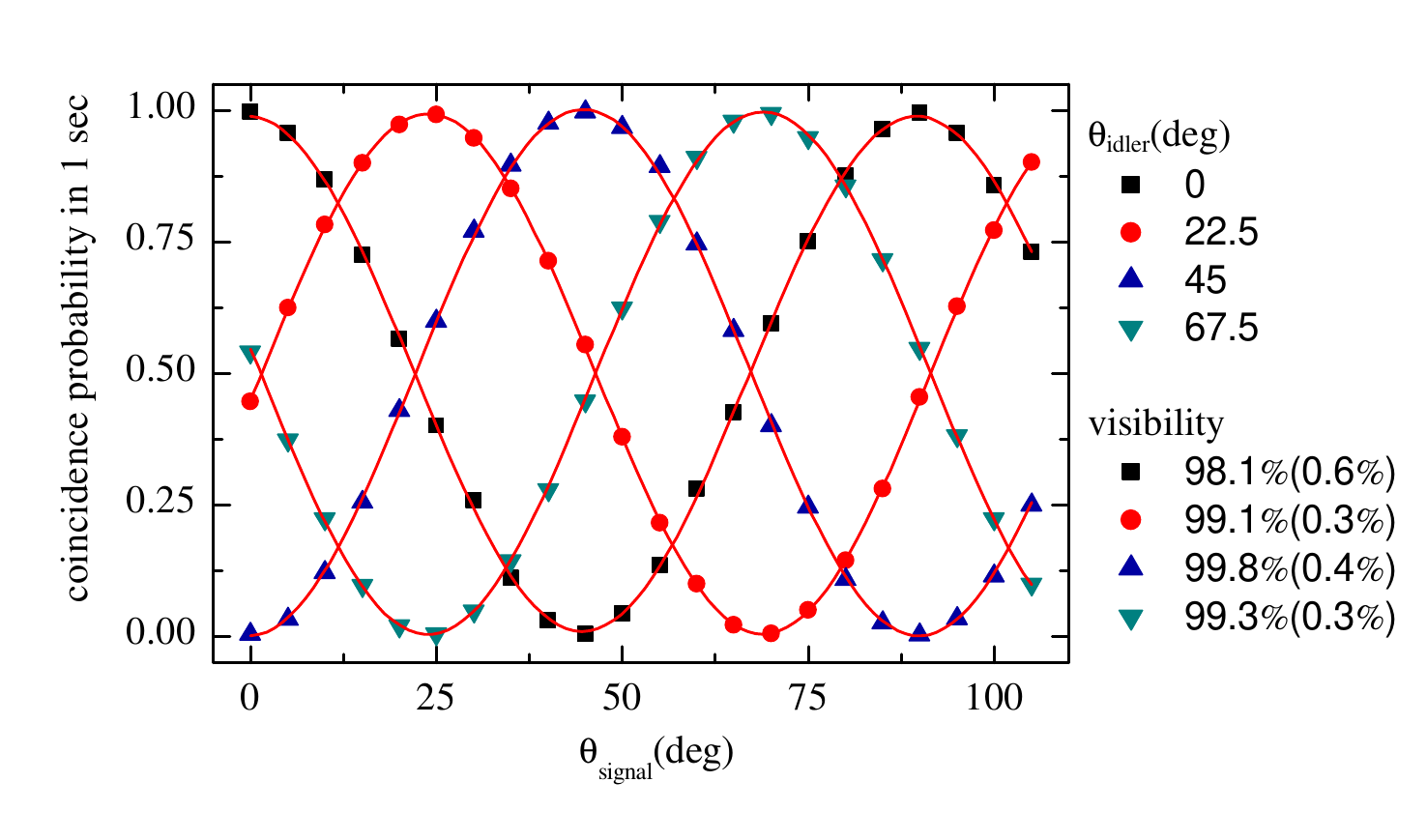}
\caption{Two-fold coincidence probability as a function of signal polarization, for four different settings of idler polarization.}
\label{fig:source}
\end{figure}

To characterize the generated entangled state, we measure the polarization correlations between signal and idler photons. We set HWPs in the signal and idler path. By setting angle of HWP, we measure a coincidence rate as a function of the two polarizers with an average number of pairs per pulse of 0.002, we obtain a high average visibility of $(99.1 \pm0.4)\%$, shown in Fig.~\ref{fig:source}.

\subsection{Bell states measurement and projection measurement}
Our Bell states measurement~(BSM) setup is able to distinguish the Bell states of $\ket{\Psi^-}=\frac{1}{\sqrt{2}}(\ket{HV}-\ket{VH})$ and $\ket{\Psi^+}=\frac{1}{\sqrt{2}}(\ket{HV}+\ket{VH})$. Photons from Client and from EPR pairs interference at the first beam splitter~(BS). The photons at state $\ket{\Psi^-}$ exit from different ports of the BS. The photons at state $\ket{\Psi^+}$ exit from the same port of the BS, and then exit from different ports of the polarization beam splitter~(PBS). We employ a BS after each port of PBS to reduce a portion of the contributions from the multi-photon pair events. The photons are detected by superconducting nanowire single photon detectors~(SNSPDs) and the detection results are analyzed by a ?eld programmable gate array~(FPGA) in real time. Furthermore, we characterize the quantum non-demolition~(QND) test by performing quantum-state-tomography measurements on the teleported quantum states. To ensure that Client and Server have a shared reference frame of polarization, we aligned rectilinear bases~(H and V) manually using fiber polarization controllers~(FPCs), and employ a phase compensator in the path of the teleported photons to compensate the difference phase.

\subsection{The interference of independent photons }
In the scheme, it requires the interference of independent photons with a high quantum-interference visibility. 
 This remains challenges for eliminating distinguishability and reducing multiple photons effect between weak coherent pulses and EPR sources. We use a three-photon Hong-Ou-Mandel interference~(HOM) to estimate the interference.

\begin{figure}
\centering
\includegraphics[width=0.45\textwidth]{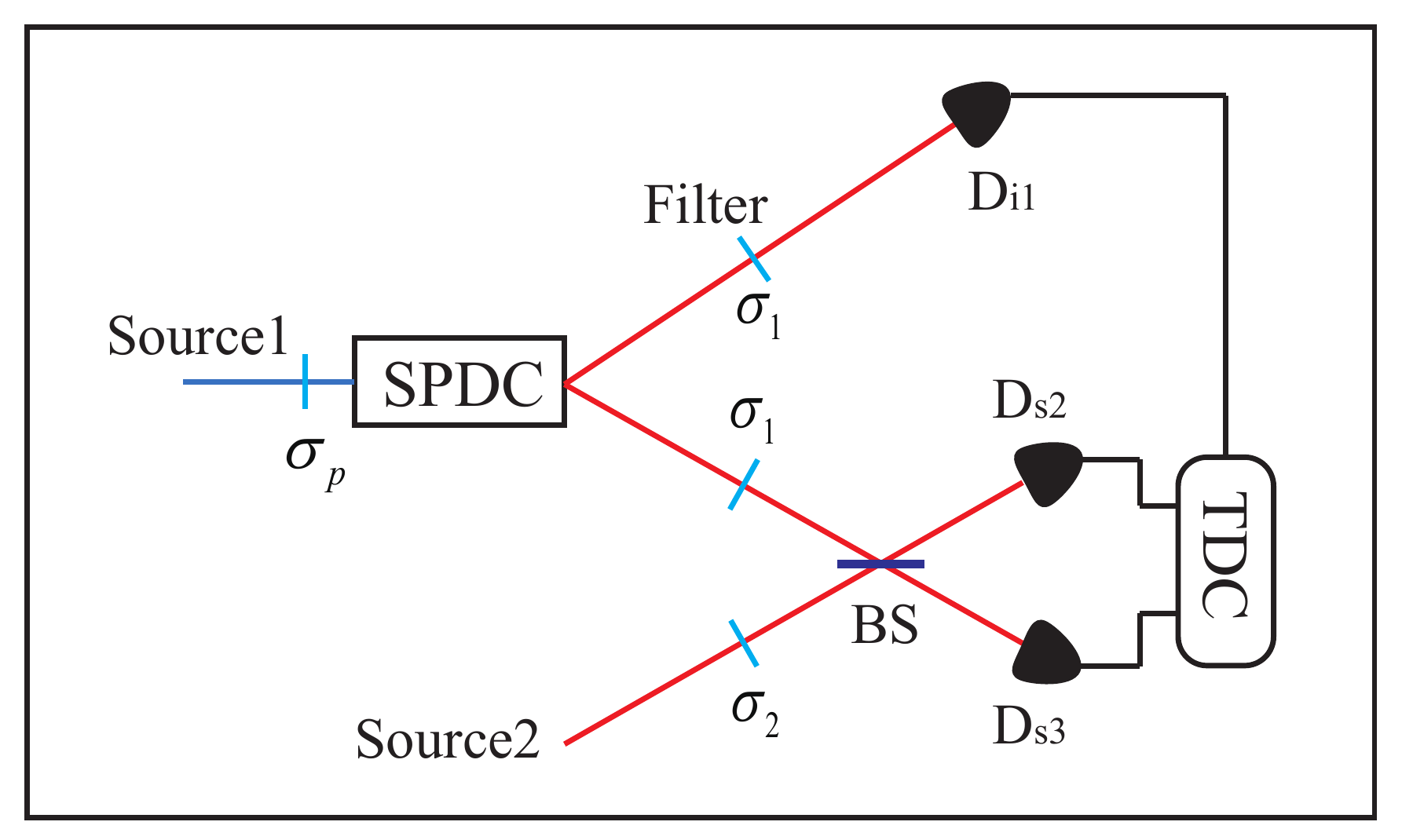}
\caption{ The Source1 is the SPDC photon pair and the Source2 is a weak coherent state. SPDC: spontaneous parametric down-conversion, BS: beam splitter, $\sigma$: the bandwidth of the filter, D: single photon detector).}
\label{fig:filter}
\end{figure}

To suppress distinguishability in spectrum, we discuss the relationship between the visibility and the bandwidth of the optical filters using the model shown in Fig.~\ref{fig:filter}. To do this, we modify the calculation in Ref. \cite{Rarity1995}. An effective 2-photon wave function at the detectors can be defined by,
\begin{equation}\label{effective}
\Psi(\bar{t_s},\bar{t_i})=\langle vac|\hat{E_s}(\bar{t}_s)\hat{E_i}(\bar{t}_i)|\Psi\rangle,
\end{equation}
where $\hat{E_{s,i}}$ are electric field operators. They are given by,
\begin{equation}\label{field}
\begin{aligned}
\hat{E_{s}}=\frac{1}{\sqrt{2\pi}}\int d\omega f_s(\omega)e^{-i\omega \overline{t}_s},
\\\hat{E_{i}}=\frac{1}{\sqrt{2\pi}}\int d\omega f_i(\omega)e^{-i\omega \overline{t}_i},
\end{aligned}
\end{equation}
here, $\hat{t}=t-\tau$ represents the moment when the photons emerge, $\tau$ is the relative delay between the inputs of the BS. Before the BS, the wavefunction is given by
\begin{equation}\label{before}
\begin{split}
\Psi_{j,k}(\bar{t_{sj}},\bar{t_{ik}})=&\alpha\int d\omega_p d\omega_{sj}d\omega_{ik} f_p(\omega_p)f_{sj}(\omega_s)f_{ik}(\omega_i)\\
&\delta(\omega_{sj}+\omega_{ik}-\omega_p)e^{i[\omega_{sj}\overline{t}_{sj}+\omega_{ik}\overline{t}_{ik}]},
\end{split}
\end{equation}
with $j=2,3$ and $k=1$. $|\alpha|^2$ is simply the probability of photon conversion in a pump pulse, the spectral functions $f_p$, $f_{sj}$ and $f_{ik}$ are limited by the filters. Then, we can express the effective wave function at detectors as,
\begin{equation}\label{wave}
\begin{split}
\Psi_{i_1,s_2,s_3}=&t^2\Psi_{12}(\overline{t}_{s12},\overline{t}_{i1})\Psi_3(\overline{t}_{s23})\\
&-r^2\Psi_{13}(\overline{t}_{s13},\overline{t}_{i1})\Psi_2(\overline{t}_{s22}),
\end{split}
\end{equation}
where $t$ and $r$ represent the transmissivity and the reflectivity of the BS. In ideal condition, $t=r=\frac{1}{\sqrt{2}}$ . We ignore the $irt$ terms as only one of detector $D_{2s}$ and $D_{3s}$ can detect the photons and the coincidence of the three detectors will be zero. Thus the probability $P(i_1,s_2,s_3)$ of detecting a threefold coincidence detection among all three detectors can be calculated from,
\begin{equation}\label{threefold}
\begin{split}
P_{i_1,s_2,s_3}=&\eta^3\int d\overline{t}_{i1}d\overline{t}_{s2}d\overline{t}_{s3}H(\overline{t}_{i1}-t_0,\Delta T)\\
&H(\overline{t}_{s2}-t_0,\Delta T)H(\overline{t}_{s3}-t_0,\Delta T)|\Psi(i_1,s_2,s_3)|^2,
\end{split}
\end{equation}
where $\eta$ is effective detector efficiencies for three detectors and $H(\overline{t}_{x}-t_0,\Delta T)$ is a normalized detector response function centered on $t_0$ that falls to zero when $\overline{t}_x-t_0>\Delta T~(\Delta T\approx1ns)$. Then with a similar procedure used in Ref.~\cite{Rarity1995}, assuming a Gaussian spectral profile, the visibility of a HOM dip is given by
\begin{equation}\label{c}
V_{HOM}=\frac{1}{\sqrt{\frac{1}{2}+\frac{\frac{\sigma_1^2}{\sigma_2^2}(\sigma_p^2+\sigma_1^2)+\frac{\sigma_1^2}{\sigma_p^2}(2\sigma_2^2+\sigma_1^2)+\frac{\sigma_p^2 \sigma_2^2}{\sigma_1^2}+3\sigma_2^2}{4(\sigma_p^2+2\sigma_1^2)}}},
\end{equation}
$\sigma_p$ is the bandwidth of the filter for the pump, $\sigma_{1}$ is for the idler, and $\sigma_{2}$ is for and signal photons and the weak coherence pulse. In our experiment, we choose $\sigma_p\approx54~pm$, $\sigma_{1}=\sigma_{2}=\sigma_{3}\approx27~pm$, thus, the computed visibility is about $98.4\%$.

To ensure a temporal overlap, we synchronize two independent sources with a microwave clock. A PPG at Client's side generates a 12.5~GHz sinusoidal signal. The signal drives an IM to modulate the continuous wave laser beam emitted by a DFB into 12.5~GHz laser pulses. The laser pulses are sent to Server and converted to an electrical signal using a 10~GHz detector. The electrical signal is amplified with a 40~GHz microwave amplifier and then used as the synchronization signal at Server's side. After synchronization, the root mean square~(RMS) value of the time jitter between the two sources is 4~ps, which is much smaller than the 133~ps coherent time of the photons.

To decrease the multi-photons effect, we optimize the mean photon number $\bar{n}_1$ of Server's EPR pairs generated per pulse and the mean photon number $\bar{n}_2$ of Client's weak coherence pulses for each fibre length. To model the interference visibility, we write the EPR state with thermal distribution,
\begin{equation}\label{c}
\varphi_1=N(\ket{00}_{s,i}+\alpha_1 \ket{11}_{s,i}+\alpha_1^2 \ket{22}_{s,i}+O(\alpha_1)),
\end{equation}
here $\mid\alpha_1\mid^2=\overline{n}_1$, $N$ is a normalising factor. The weak coherent state can be wrote with Poissonian distribution,
\begin{equation}\label{c}
\varphi_2=e^{-\frac{|\alpha_2|^2}{2}}(\ket{0}+\alpha_2 \ket{1}+\frac{\alpha_2^2}{\sqrt{2!}} \ket{2}+O(\alpha_2)),
\end{equation}
here $\mid\alpha_2\mid^2=\overline{n}_2$. Then we follow a similar procedure as in Ref. \cite{Fulconis2007}. Considering the probability of triple coincidence, the visibility is given by
\begin{equation}\label{c}
\begin{split}
&V_{HOM}\approx\\
&\frac{4\bar{n}_2+2\bar{n}_2^2(2-\eta)+8\bar{n}_1(1-\frac{\eta_i}{2})\bar{n}_2(2-\eta)}{4\bar{n}_2+3\bar{n}_2^2(2-\eta)+12\bar{n}_1(1-\frac{\eta_i}{2})\bar{n}_2(2-\eta)+8\bar{n}_1},
\end{split}
\end{equation}
here $\eta$ is the detection efficiency for HOM, $\eta_i$ is the detection efficiency for the idler photon. Considering we reduce the contributions from the multi-photon-pair events, the visibility can be wrote as,
\begin{equation}\label{d}
\begin{split}
&V_{HOM}\approx\\
&\frac{2\bar{n}_2+\frac{\bar{n}_2^2}{2}(2-\eta)+2\bar{n}_1(1-\frac{\eta_i}{2})\bar{n}_2(2-\eta)}{2\bar{n}_2+\frac{3\bar{n}_2^2}{4}(2-\eta)+3\bar{n}_1(1-\frac{\eta_i}{2})\bar{n}_2(2-\eta)+4\bar{n}_1},
\end{split}
\end{equation}

\begin{figure}
\centering
\includegraphics[width=0.45\textwidth]{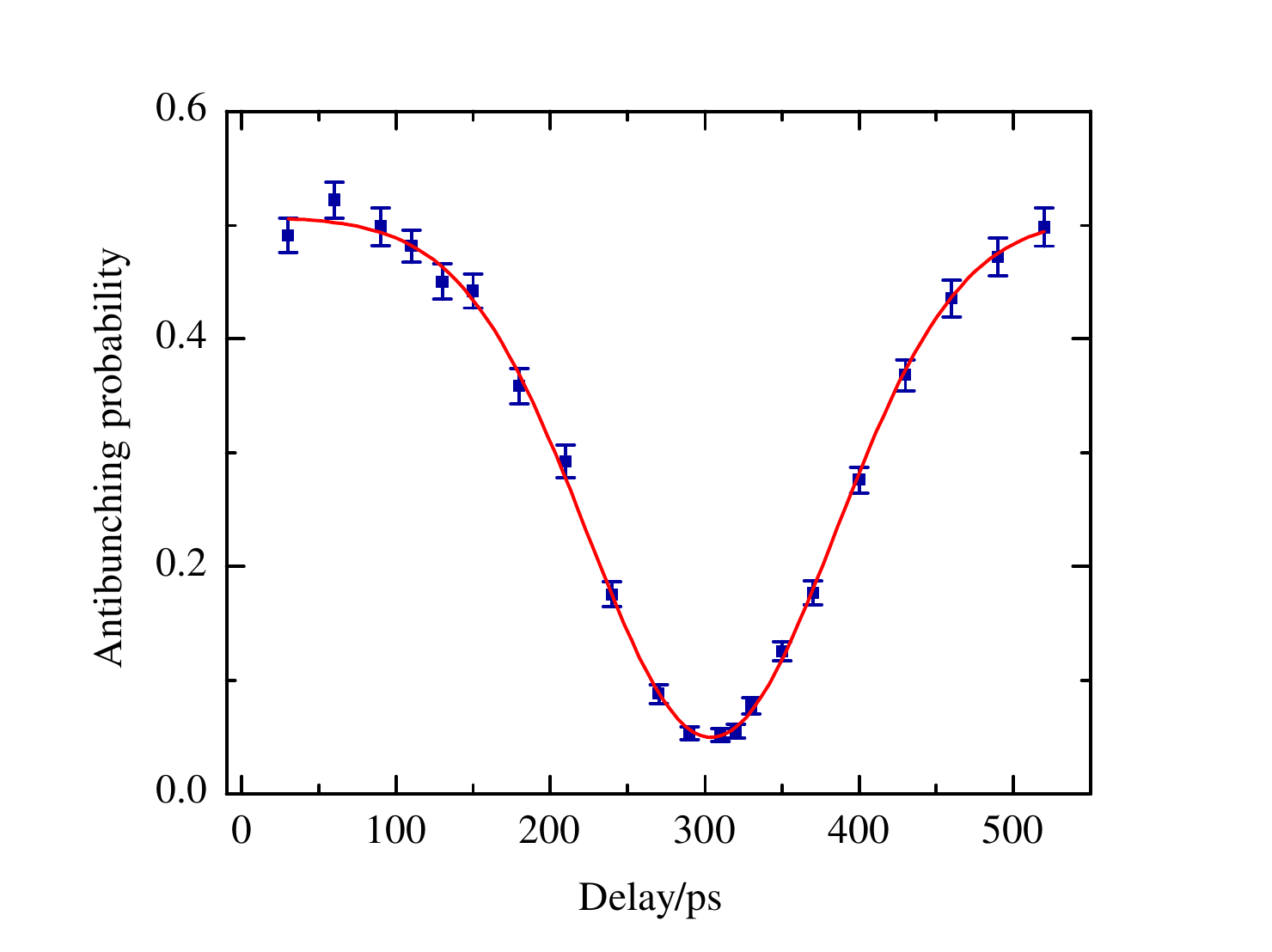}
\caption{\textbf{Experiment result of HOM interference.} The visibility of the fitted curve is $V_{HOM}=(90.2\pm0.4)\%$.}
\label{fig:HOM}
\end{figure}

In addition, the single-mode fiber ensure the indistinguishability in the spatial degree of freedom. In our experiment, $\eta\approx0.105$, $\eta_i\approx0.08$, we optimize the mean photon number per pulsed for each distance, take 0~km as an example, we set $\bar{n}_1=0.0020$ and $\bar{n}_2=0.0645$ , consider both spectrum distinguishability and multiple photons effect, in theory $V_{HOM}=92.5\%$, and we get experimental result with $V_{HOM}=(90.2\pm0.4)\%$, shown in Fig.~\ref{fig:HOM}.

\section{I1DC subroutine simulation} \label{app:I1DC}
To generate blind qubits, Server has to conduct the I1DC subroutine~\cite{Dunjko2012b} after Client telling him where the signal pulses are. However before he performing the operation, the qubits need to be stored in quantum memory, which is beyond our power. So we gather information about the states received by Server, and then simulate this subroutine in a classical way.
The I1DC subroutine runs as follows:
Assume there are n signal qubits left in Server's side. Devide the n signal qubits into S groups, each group contains k qubits. For the k qubits:

1. For i=1 to k-1

	(a) Apply the unitary $ctrl-Z(H\otimes I)$ to qubits i and i+1.

	(b) Measure qubit i in the Pauli-X basis, get the outcome $y_i$.

2. Report the measurement results $y=(y_1, ..., y_{(k-1)})$ and the remaining qubit i=k in state $\ket{+_{\theta_k}}$.

Through tomography, we obtain the density matrixes $(\rho_1, \rho_2, ...)$ of the signal states at Server's side. With these density matrixes we can conduct the I1DC simulation. In the experiment for 0~km, we get 4384 signal pulses, which means for 1000 groups, each group will contain 4 or 5 pulses. Applying I1DC to each group, we will get 3 or 4 measurement results $y=(y_1, y_2, ...)$ and a density matrix $\rho_{experiment}$ of the remaining qubit. Furthermore, we compare the remaining qubit $\rho_{experiment}$ with the idea state $\rho_{idea}$ by calculating the fidelity. Here $\rho_{idea}$ is obtained by assuming a perfect QND measurement. Finally, we get fidelity of $(81.94\pm1.95)\%$.

\section{Details of experimental results} \label{app:results}
In our experiment, we have run the system at different distances of 0 km, 26 km, 50 km, 76 km and 100 km for preparing 1000 blind single qubits. Considering both the QND fidelity and the required number N, we optimized the signal and decoy states intensities for each distance. Details of results are listed in Tables~\ref{tab:res}. The gains are obtained from the BSM results, and the  lower bound of the required number N are calculated by using Eq.~(9). Furthermore, the fidelities of the quantum states after QND measurement are calculated with the reconstructed density matrices via relation $F_{\theta_i}=\langle\Phi_{QND}|\Phi_{\theta_i}|\Phi_{QND}\rangle$ shown in Tables~\ref{tab:fidelityp}. The uncertainties are calculated using a Monte Carlo routine assuming Poissonian errors.

\begin {table*}
\centering
\caption{Details of the experimental results for $S=1000$ for various distances L. N is the lower bound of required pulse number, $Q_\mu$, $Q_\nu$ and $Q_0$ are the gains for the signal states, decoy states  and vacuum states, respectively. The error bars represent one standard deviation.}
\vspace{20pt}
\resizebox{\textwidth}{18mm}{
\begin{tabular}{ccccc}

\hline
\hline
 L~(km)&$N$&$Q_\mu$&$Q_\nu$&$Q_0$\\
\hline
0&$7.8769\times10^{10}$&$(1.1286\pm0.0003)\times10^{-5}$&$(6.1794\pm0.0034)\times10^{-7}$&$(1.7951\pm0.0034)\times10^{-8}$\\
26&$2.1564\times10^{11}$&$(1.1312\pm0.0003)\times10^{-5}$&$(1.3192\pm0.0012)\times10^{-7}$&$(1.4389\pm0.0031)\times10^{-8}$\\
50&$4.5295\times10^{11}$&$(1.0666\pm0.0002)\times10^{-5}$&$(4.1489\pm0.0067)\times10^{-8}$&$(1.7397\pm0.0025)\times10^{-8}$\\
76&$1.0822\times10^{12}$&$(1.5522\pm0.0009)\times10^{-6}$&$(2.5001\pm0.0052)\times10^{-8}$&$(1.2888\pm0.0022)\times10^{-8}$\\
100&$1.8853\times10^{12}$&$(5.8679\pm0.0042)\times10^{-7}$&$(2.9827\pm0.0041)\times10^{-8}$&$(1.2303\pm0.0015)\times10^{-8}$\\
\hline
\hline
\label{tab:res}
\end{tabular}}
\end{table*}
\begin {table*}
\centering
\caption{The reconstructed density matrices for eight polarization states $\theta _i\in_R \{k\pi\backslash 4 \colon 0\leq k\leq7 \}$ after QND measurement over each distance L.}
\vspace{20pt}
\resizebox{\textwidth}{15mm}{
\begin{tabular}{ccccccccc}

\hline
\hline
 L~(km)&$k=0$&$k=1$&$k=2$&$k=3$&$k=4$&$k=5$&$k=6$&$k=7$\\
\hline
0&$(93.9\pm1.3)\%$&$(93.5\pm1.1)\%$&$(96.0\pm1.2)\%$&$(95.7\pm0.7)\%$&$(95.8\pm1.3)\%$&$(95.1\pm0.9)\%$&$(95.9\pm1.3)\%$&$(95.2\pm0.8)\%$\\
26&$(94.1\pm1.4)\%$&$(93.3\pm0.9)\%$&$(95.1\pm1.3)\%$&$(93.8\pm1.1)\%$&$(93.0\pm1.4)\%$&$(93.7\pm1.0)\%$&$(96.6\pm1.3)\%$&$(94.5\pm0.8)\%$\\
50&$(91.4\pm0.8)\%$&$(95.4\pm0.9)\%$&$(92.2\pm0.7)\%$&$(91.2\pm1.1)\%$&$(90.8\pm0.8)\%$&$(94.5\pm1.0)\%$&$(91.5\pm0.7)\%$&$(90.0\pm1.3)\%$\\
76&$(91.7\pm1.6)\%$&$(90.4\pm2.4)\%$&$(92.9\pm1.1)\%$&$(92.7\pm2.2)\%$&$(92.4\pm1.4)\%$&$(90.2\pm2.0)\%$&$(91.8\pm1.4)\%$&$(91.0\pm1.9)\%$\\
100&$(85.8\pm1.3)\%$&$(86.8\pm1.8)\%$&$(86.9\pm1.5)\%$&$(86.4\pm1.7)\%$&$(85.9\pm1.3)\%$&$(88.7\pm1.8)\%$&$(86.9\pm1.3)\%$&$(87.7\pm1.6)\%$\\
\hline
\hline

\end{tabular}}
\label{tab:fidelityp}
\end{table*}

%

\end{document}